\documentclass{aastex6}
\usepackage{natbib}

\newcommand{\be}           {\begin{equation}}
\newcommand{\ee}           {\end{equation}}
\newcommand{\bea}          {\begin{eqnarray}}
\newcommand{\eea}          {\end{eqnarray}}
\newcommand{\WISE}       {{\sl WISE}}

\slugcomment{submitted to the Astronomical Journal}

\shorttitle{Albedo distribution}
\shortauthors{Wright et al.}

\begin{document}

\title{The Albedo Distribution of Near Earth Asteroids}

\author{Edward L.\ Wright\altaffilmark{1}}
\affil{UCLA Astronomy, PO Box 951547, Los Angeles CA 90095-1547}
\author{Amy Mainzer\altaffilmark{2}}
\author{Joseph Masiero\altaffilmark{2}}
\author{Tommy Grav\altaffilmark{3}}
\and
\author{James Bauer\altaffilmark{2,4}}

\altaffiltext{1}{wright@astro.ucla.edu}
\altaffiltext{2}{Jet Propulsion Laboratory, California Institute of Technology, 
4800 Oak Grove Dr., Pasadena, CA, 91109-8001, USA}
\altaffiltext{3}{Planetary Sciences Institute, 1700 E Fort Lowell Rd \#106, Tucson, AZ 85719}
\altaffiltext{4}{Infrared Processing and Analysis Center, 770 South Wilson Ave., Pasadena, CA 91125}

\begin{abstract}

The cryogenic WISE mission in 2010 was extremely sensitive to
asteroids and not biased against detecting dark objects.  The albedos
of 428 Near Earth Asteroids (NEAs) observed by WISE during its fully
cryogenic mission can be fit quite well by a 3 parameter function
that is the sum of two Rayleigh distributions.  
The Rayleigh distribution is zero for negative values, and follows $f(x) = x
\exp[-x^2/(2\sigma^2)]/\sigma^2$ for positive $x$.  The peak value
is at $x=\sigma$, so the position and width are tied together.  
The three parameters are the fraction of the objects in the dark
population, the position of the dark peak, and the position of the
brighter peak.  We find that 25.3\% of the NEAs observed by WISE are
in a very dark population peaking at $p_V = 0.030$, while the other
74.7\% of the NEAs seen by WISE are in a moderately dark population
peaking at $p_V = 0.168$.  A consequence of this bimodal distribution
is that the Congressional mandate to find 90\% of all NEAs larger
than 140 m diameter cannot be satisfied by surveying to H=22 mag,
since a 140 m diameter asteroid at the very dark peak has H=23.7
mag, and more than 10\% of NEAs are darker than $p_V = 0.03$.

\end{abstract}


\section{Introduction}

The Wide-field Infrared Survey Explorer (\WISE) \citep{wright/etal:2010}
mapped the entire sky between 14 Jan 2010 and 17 Jul 2010, then
continued on to map the entire sky again prior to 1 Feb 2011.  On
7 Aug 2010 the outer cryogen tank ran out of solid hydrogen coolant,
ending the 4 band cryogenic phase of the WISE mission.
The NEOWISE mission is a separately funded program to search
for Near Earth Objects (NEOs) in the WISE data.  NEOs include both
asteroids and comets.  In this paper we will only consider NEAs.
In this paper we study the sample of
428 NEAs observed with all 4 WISE bands during 2010 \citep{mainzer/etal:2011c}.  
The wide range of wavelengths
spanning the peak of the thermal infrared emission allows the Near Earth Asteroid
Thermal Model (NEATM, \cite{harris:1998}) to work very well, 
since the
4.6, 12  and 22 $\mu$m fluxes are all dominated by thermal emission for NEAs 
and their ratios provide a tight constraint on the beaming parameter $\eta$,
while the 22 $\mu$m flux gives a diameter that is only weakly dependent
on $\eta$.  This NEOWISE sample
of NEAs with radiometric diameters is more than an order of magnitude larger
than the sample of 36 objects studied by \cite{stuart/binzel:2004}.  We find in this
paper that a simple 3 parameter model provides a very useful approximation
to the observed distribution of albedos.

\citet{masiero/etal:2011} found bimodal albedo
distributions for the the inner, middle and outer main belt asteroid populations
with different dark fractions as a function of distance from the Sun.
The optical colors of asteroids are known to correlate with albedo, with the 
higher albedo S (``stony'') type being redder than the lower albedo C (``carbonaceous'')
type \citep{bowell/lumme:1979}. 
Thus the albedo distribution of the NEAs could give clues about the source or sources of the
NEA population.
Indeed, \citet{granvik/etal:2016} find that the high albedo and low albedo fractions of
the NEOWISE NEA sample have significantly different parent distributions, confirming
a result from \citet{mainzer/etal:2012}.
They also find that there are fewer observed NEAs with small
perihelia than are predicted by their models, 
and suggest that the low perihelion ($q < 0.2$ AU) objects have been thermally disrupted, 
based on a large sample of NEAs from the Catalina Sky Survey.
They suggest that the darker objects are more subject to disruption by thermal
stress when close to the Sun, but there are only 5 low perihelion ($q < 0.2$ AU) NEAs
in the NEOWISE sample, so this last suggestion requires a larger sample of NEA albedos 
for verification.

The NEA albedo distribution also enters into estimates of the hazard due to
Earth impacts.
In 2005, Congress gave NASA the goal of finding 90\% of all Near Earth Objects
larger than 140 meters in diameter\footnote[5]{National Aeronautics and Space 
Administration Authorization Act of 2005 (Public Law 109-155), January 4, 2005, 
Section 321, George E. Brown, Jr. Near-Earth Object Survey Act.}.  
One use of the albedo distribution proposed here
is to determine what optical limiting magnitude is needed to meet this objective.
The optical absolute magnitude is given by 
$H = 5\log([1329\;\mathrm{km}]/D) - 2.5\log(p_V)$ \citep{bowell/etal:1989}, 
and the usual assumption that
$H=22$ mag corresponds to $D = 140$~m requires that $p_V = 0.142$.  But since
$p_V$ is distributed over a wide range of values, the actual optical limiting
magnitude needed to satisfy the mandate
depends on the size distribution of NEAs, with shallower slopes leading
to somewhat relaxed search limits.  

\section{Observations}

The dataset used is the collection of 428 NEAs observed by WISE during the
fully cryogenic phase of the mission: the seven months from January 7 to
August 7, 2010.  This is the same data used by \cite{mainzer/etal:2011c}.
Of the 428 NEAs in the dataset, 9 have only WISE observations
with no optical followup.  These objects have an unknown albedo,
and are not used in the fit.  They are perhaps
preferentially low albedo objects, but this is only a potential 2\% bias in
the abundance of dark objects.  However, these objects have fairly long
arcs in the WISE data.  Many other WISE tracklets are shorter arcs, and if
these receive no optical followup, they are filed by the Minor Planet Center
(MPC) without an orbit.
\citet{mainzer/etal:2011c} noted that during the 4 band fully cryogenic portion
of the NEOWISE mission about 15-20 NEO candidates appeared
on the MPC NEO Confirmation Page but received no followup.

Thus while using infrared observations eliminates the discovery bias against
dark objects, there can still be a followup bias against dark objects.  An example
is 2015 SS$_{20}$, which was on the MPC NEO Confirmation Page
for two weeks without receiving any optical followup.  It was designated
2015 SS$_{20}$ and filed without an orbit.  Searching the NEOWISE image
data for $3\sigma$ bumps led to a tentative longer observational arc, which led
to a very faint counterpart on previously obtained CHFT MegaCam frames
\citep{forshay/etal:2015}, 
and then to a cross identification with 2015 WL$_{16}$, another NEOWISE tracklet
that was also designated without followup or an orbit.  
IR fits to the NEOWISE data show that the diameter
is definitely larger than 140 m, but the H magnitude is 22.7 in \cite{forshay/etal:2015}.
While it is tempting to assume that the low albedo 
of this object led to the lack of followup,
it is quite likely that the phase of Moon,  
which was waxing gibbous at the discovery 
2015 SS$_{20}$ and full at the discovery of 2015 WL$_{16}$ had a large
effect as well.
We have used this example to inform our assumption about the nature
of the 9 objects that have only WISE observations.  Three of these objects were discovered
in a three day interval just before the full Moon in June 2010.  Another three were
discovered at extreme southern declinations $\delta < -72^\circ$, where follow-up 
opportunities were limited.  
As a result we have assumed that these 9 objects were 
missed for reasons other than a very low albedo, and have not made any correction
to the dark fraction derived from the 419 objects that were followed up by optical
observers.

\section{Previous Fits}

\cite{mainzer/etal:2011c} used a 5 parameter double Gaussian to fit the NEA albedo
distribution:
\be
p_{2G}(p_V)  = f \exp\left(-\frac{(p_V-d)^2}{2e^2}\right)
 +  
c \exp\left(-\frac{(p_V-b)^2}{2a^2}\right)
\label{eq:2G}
\ee
While this appears to have 6 parameters, one of the degrees of freedom is
taken away by the normalization constraint, $\int p(p_V)dp_V = 1$.  
The dark albedo peak was
found to be $d = 3.4$\%, while the bright albedo peak was found to
be $b = 15.1$\%.   
This two Gaussian model for the albedo distribution has the conceptual
problem that it predicts a non-zero probability density for small negative
albedos, while the albedo is actually constrained to be non-negative.
This is easily solved by replacing the non-zero density by zero.
A related problem is that the two Gaussian model overpredicts the
abundance of very low albedos, because the probability density does
not go to zero as the albedo goes to zero.

Note that the dark Gaussian accounts for 28.8\% of the
cumulative probability in this two Gaussian model.

\section{Rayleigh Distribution Fits}

The problems with the two Gaussian model can be fixed by
replacing the Gaussian by a Rayleigh distribution.  The Rayleigh
distribution is the distribution of the radius in a two dimensional
Gaussian.  It is clearly zero for negative values, because the radius
is always non-negative.  The probability density goes to zero as
the radius goes to zero.  The full formula for a Rayleigh distribution
is
\be
p(x) = \frac{x \exp(-x^2/2\sigma^2)}{\sigma^2}.
\ee
% y = x^2/2\sigma^2, dy = xdx/\sigma^2
% n^th moment: x^n = \sigma^n (2y)^{n/2}, so \sigma^n 2^{n/2} \Gamma(n/2+1)
% so mean x is \sigma \sqrt{2} \sqrt{pi}/2 = \sqrt{pi/2}\sigma
% so <x^2> = 2\sigma^2, \var{x} = \sigma^2 (2-pi/2) 
The mean of $x$ is $\langle x \rangle = \sigma \sqrt{\pi/2}$, and
the fractional width of the distribution is 
$\sqrt{\mathrm{var}(x)}/\langle x \rangle = 0.5227$.
This ratio can be compared to $d/e = 0.41$ and $b/a = 0.81$
in the two Gaussian model of Eqn(\ref{eq:2G}).
The peak of $p(x)$ occurs at $\sigma$, but the most likely value
of $\ln(x)$ occurs at $\sigma\sqrt{2}$.

The bimodality of the albedo distribution requires that the full model
include two Rayleigh distributions, giving the formula
\be
p_{2R}(p_V) = f_D \frac{p_V e^{-p_V^2/2d^2}}{d^2}
+ (1-f_D) \frac{p_V e^{-p_V^2/2b^2}}{b^2}
\label{eq:2R}
\ee
This is an example of a finite mixture model in statistics
\citep{mclachlan/peel:2000} and the particular case of two Rayleigh distributions
has been used in a very different application to failure time distributions
by \citet{attia:1993}.

\begin{figure}[tb]
\plotone{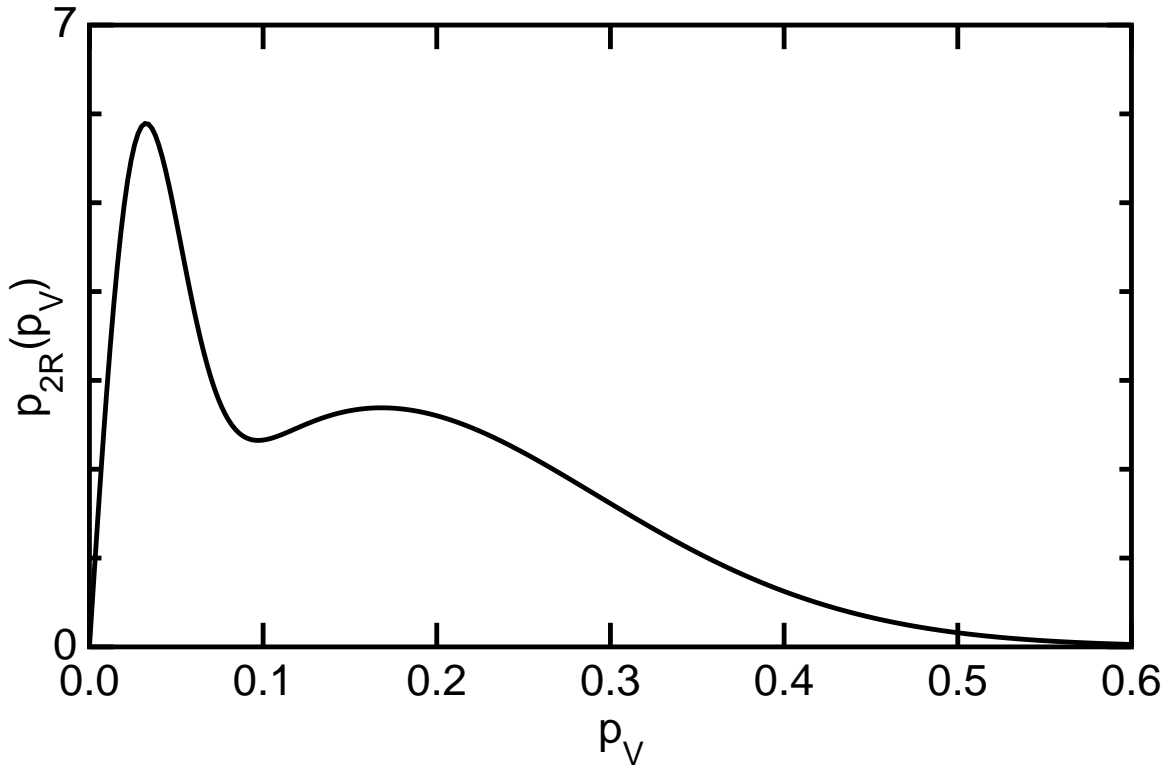}
\caption{The two Rayleigh distribution model for the probability
density function of Near Earth Asteroid albedos.\label{fig:pdf}}
\end{figure}

\begin{figure}[tb]
\plotone{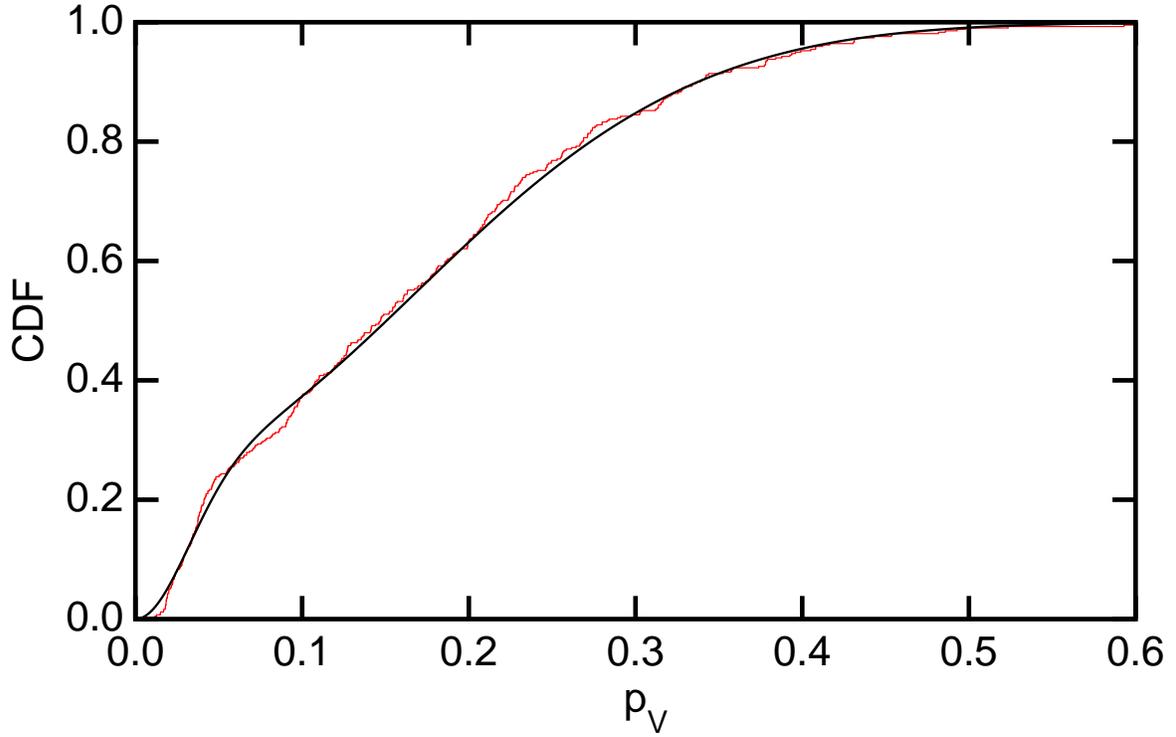}
\caption{Comparison of the cumulative distribution function (CDF) for
the observed set of NEAs (red staircase function) with the CDF for
the two Rayleigh distribution model.\label{fig:KS}}
\end{figure}

\begin{figure}[tb]
\plotone{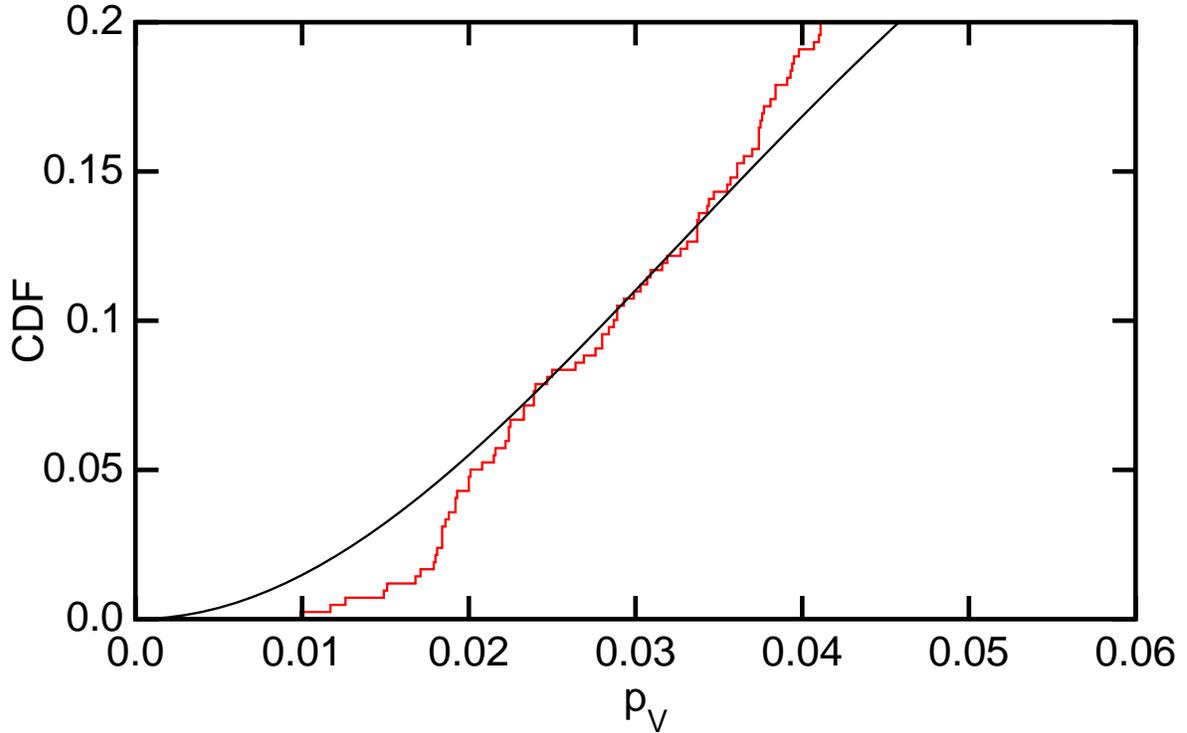}
\caption{The low albedo corner of Figure \ref{fig:KS}.
\label{fig:KS-inset}}
\end{figure}

The three parameters in Eqn(\ref{eq:2R}) were adjusted to
minimize the maximum deviation between the observed
and model cumulative distribution functions.  This corresponds
to minimizing the Kolmogorov-Smirnov statistic.  The
motivation for using this criterion to optimize the parameters
comes from the congressional goal: we want to have a model
that is very close to reality at the $10^{th}$ \%-tile of the distribution
in order to be sure to find 90\% of all NEAs bigger than 140 meter
diameter.   The resulting parameters are the dark fraction
$f_D = 0.253$, the dark peak $d = 0.030$, and the bright peak
$b = 0.168$.  The best fit $p(p_V)$ is shown in Figure \ref{fig:pdf}.
The goodness of fit is shown in Figures \ref{fig:KS} \&
\ref{fig:KS-inset}.

The best fit has a maximum deviation between the model and
observed CDFs of $\Delta = 0.027$.
If we had not adjusted the 3 parameters for a best fit,
the probability of a deviation larger than this due to chance
fluctuations alone in the Kolmogorov-Smirnov test
would be 92\%.  But adjusting parameters
to minimize $\Delta$ has a strong effect on the probabilities.
To evaluate this effect we generated random datasets
using the best fit model and the methods in \S\ref{sec:applications}, 
and then readjusted the three parameters for
a best fit to each random dataset.  
We got better fits with lower $\Delta$ than the observed data
for 96\% of the random datasets.  Thus the two Rayleigh distribution
model is a convenient and acceptable fit, not ruled out by the current data, 
but should not be taken as a final description of
the albedo distribution.   The scatter in the parameters for the fits to the
random datasets were $\sigma(f_D) = 0.032$, $\sigma(d) = 0.003$,
and $\sigma(b) = 0.006$.
%    964/1000     f_d= 0.254 +/- 0.032    d= 0.030 +/- 0.003    b= 0.168 +/- 0.006
The dark fraction is determined to a relative accuracy of 13\% by
the NEOWISE dataset and the dark albedo peak is determined to a relative
accuracy of 10\%.

\section{Main Belt Albedo Distribution}

\citet{masiero/etal:2011} fit mixture models with two log normal distributions to
the main belt asteroid albedos derived from NEATM fits to the WISE 4-band data.
This dataset contains over $10^5$ objects.  Thus \citet{masiero/etal:2011}  were able
to subdivide the dataset into inner, middle and outer main belt samples, and fit for
a five parameter dual log normal distribution in each subsample.

The three main belt albedo distributions all have a dark albedo peak at 
$p_V = 0.06$ with a width shown by $\sigma_+ = 0.03$ and $\sigma_- = 0.02$.
Thus the $\sigma$ in the logarithm was $\ln(1.5) = 0.41$.   
The most likely logarithm of $p_V$ in the dark peak of the two Rayleigh distribution in
Eqn(\ref{eq:2R}) occurs at $b\sqrt{2} = 0.042$ compared to $0.06$ in the main belt, 
so the NEA albedo distribution has
a darker and wider dark peak than the main belt distributions.
The fraction of objects
in the dark peak was 47\% to 73\%, so the MBA
albedo distributions in \citet{masiero/etal:2011}  have  a larger fraction of
dark asteroids than the NEA distribution.

\section{Cumulative Detection Limits}

\begin{figure}[tb]
\plotone{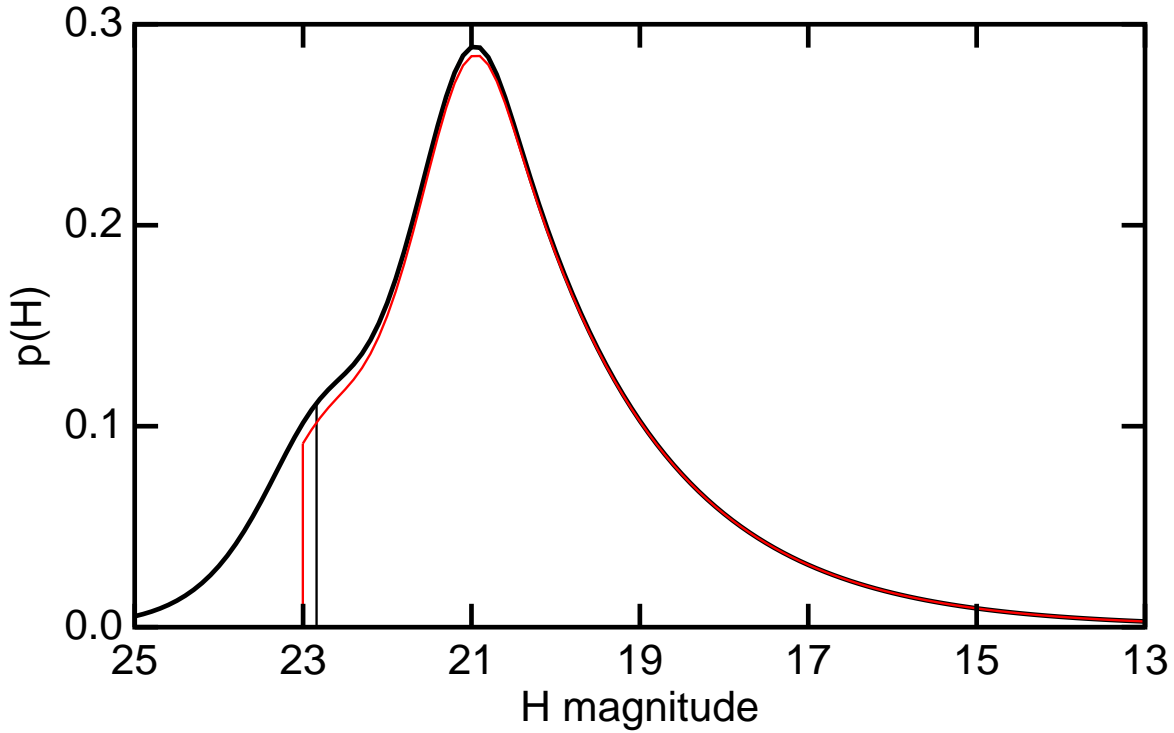}
\caption{The thick curve shows the distribution of $H$ magnitudes for
asteroids with diameters greater than 140 m and $N(>D) \propto D^{-1.3}$.
The thin red curve shows this $p(H)$ multiplied by a simple
survey completeness model with 90\% completeness at $H=23$ mag
and 96\% completeness at $H=22$ mag.  90\% of the $D > 140$ m asteroid
population is contained in the area under the thin red curve
to the right of vertical red line at $H=23$.
A survey that is 100\% complete for $H$ brighter than the thin vertical
black line at $H = 22.84$ mag will also find 90\% of objects bigger that 140 m.
\label{fig:pH}}
\end{figure}

The albedo distribution derived here has 90\% of NEAs with $p_V > 0.03$.
Thus if one wants to find 90\% of all 140 m sized NEAs then one must
search down to $H = 23.7$.  But the Congressional mandate is to
find 90\% of all NEAs 140 meters or larger, and finding the necessary
$H$ mag limit for this criterion depends on the size distribution of NEAs. 
For example, if the size distribution is very shallow, most of the objects larger
than 140 m diameter will be much larger than 140 m and thus easier to find.
In this paper we will use $N(>D) \propto D^{-1.3}$ which is the slope
found by \cite{ivezic/etal:2002} for main belt asteroids 
with $D < 5$~km using the Sloan
Digital Sky Survey.  The slope also agrees with 
$N(>D) \propto D^{-1.32 \pm 0.14}$ found for NEAs between 100 meters
and 1 km  by 
\cite{mainzer/etal:2011c}.
The probability density for  H magnitudes 
for a given size distribution $p(D)$ with $D > D_m$ and
albedo distribution $p(p_V)$ is
\be
p(H)  =  \int_{D_m}^\infty
\int_0^\infty \delta\left(H-5\log\left(\frac{1329\;\mathrm{km}}{D \sqrt{p_V}}\right)\right) 
p(p_V) d p_V p(D) dD
\label{eq:pH}
\ee
when we assume that the albedo and size are independent, since
\cite{mainzer/etal:2011c} did not see a correlation of
albedo and size.
Note that geometric albedos like $p_V$ can be larger than 1 if
the phase function is sharply peaked:
Scotchlite\textregistered  ~tape is an example.
But in any case these high values of $p_V$ contribute little
to the integral.
For any $D$, the value of $p_V$ that gives $H$ is
\be
p_V = \left(\frac{1329\;\mathrm{km}}{D}\right)^2 10^{-0.4 H}.
\ee
The derivative $d p_V/dH$ needed to evaluate Eqn(\ref{eq:pH})
is given by $d p_V = -(0.4 \ln 10) p_V dH$.
Inserting these values for $p_V$ and $d p_V$ into Eqn(\ref{eq:pH})
gives
\be
p(H)  =   (0.4 \ln 10) \int_{D_m}^\infty p_{2R}(p_V) p_V p(D) dD
\ee
which is plotted in Figure \ref{fig:pH}.  The vertical black line
in this figure is at $H = 22.84$ mag, and the area to the right of this
line is 90\% of the total.  So an optical survey that is 100\% complete
for $H < 22.84$ will detect 90\% of all asteroids larger than 140 m
diameter for $N(>D) \propto D^{-1.3}$.  Similar limits for the
main belt albedo distributions in \citet{masiero/etal:2011} are
$H < 22.5$, $<22.6$ and $<22.8$ mag for the inner, middle and outer
main belt subsamples.

But 100\% completeness is not realistic.  The red curve in
Figure \ref{fig:pH} shows $p(H)$ multiplied by an completeness
function that is 90\% for $H = 23$ mag and with the incompleteness scaling
inversely with the flux, or $0.1 \times 10^{0.4(H-23)}$.  Thus the
completeness is 96\% at $H=22$ and 99\% at $H=20.5$.  The red
vertical line is at $H=23$ mag, and the area to the right of this line contains
90\% of all the asteroids with diameter greater than 140 m.

The distribution of $H$ in Figure \ref{fig:pH} is very similar to Figure
6 of \citet{grav/mainzer/spahr:2016} who used Monte Carlo methods
with albedos chosen randomly from the actual list of albedos in
\citet{mainzer/etal:2011c}.  Thus the conclusion that optical surveys
should strive for completeness down to $H = 23$ mag is not affected
by the details of the fit.

\section{Useful Applications}
\label{sec:applications}

One potential use for the albedo distribution shown here is in
simulations of surveys.  For this one needs to randomly choose
$p_V$ from the distribution in Eqn(\ref{eq:2R}).  This is easily
accomplished using two independent random variates $x$ and $y$
drawn from a uniform distribution over $[0,1)$.  The procedure
takes $t = d$ if $x < f_D$ or $t=b$ otherwise, and then
$p_V = t\sqrt{-2\ln(1-y)}$.

A second use for Eqn(\ref{eq:2R}) is as  a Bayesian prior on the
albedo when fitting for the diameter of an NEA based on limited
infrared data such only 3.4 and 4.6 $\mu$m.   This allows reasonable 
corrections for the effects of scattered sunlight on the 3.4 $\mu$m flux of an 
NEA even if no optical data are available.  With no
optical data a higher albedo 
means less of the 3.4 $\mu$m flux is thermal, so the asteroid
must be cooler, leading to a higher $\eta$ and a larger diameter.
This variation of diameter with albedo is smaller than 
the $D \propto {p_V}^{-0.5}$ when only optical data are available,
and it has the opposite sense.
For example, if an NEA 1.155 AU from the Sun with $\eta=1$, 
emissivity $\epsilon=0.9$, albedo $p = 0.168$ 
and slope parameter $G= 0.15$ is observed at 60$^\circ$ phase angle,
but is analyzed assuming an albedo $p = 0.03$, the NEATM derived diameter based on
3.4 and 4.6 $\mu$m data alone goes down
by a factor of 0.83, while the diameter derived from optical data alone goes up
by a factor of $\sqrt{0.168/0.03} = 2.37$. 
This degeneracy between albedo and diameter in optical-only
or infrared-only cases
can be constrained by using an informative prior on the albedo.
A Monte Carlo Markov Chain using Eqn(\ref{eq:2R}) as
a prior can give a good picture of the resulting diameter uncertainty
that allows for the non-Gaussianity of the albedo distribution.
This prior can be implemented by using a uniform prior on
$\ln(p_V)$ and applying a penalty to $\chi^2$ of
$-2\ln[p_V p_{2R}(p_V)]$. 

\section{Conclusions}

The distribution of albedos for Near Earth Asteroids is very broad,
and it can be written as the sum of two Rayleigh distributions
with peaks that differ by a factor close to 6 in albedo.  
The existence of the dark peak in the albedo distribution is very well 
established by the NEOWISE dataset, with a confidence of 8$\sigma$.
As a result, the Congressional goal to find 90\% of all NEAs larger
than 140 m diameter can not be met by surveying to a limit of $H < 22$ mag.  
Optical surveys must aim for very substantial completeness
down to $H = 23$ mag to satisfy the mandate. 

\acknowledgments

This publication makes use of data products from the Wide-field
Infrared Survey Explorer, which is a joint project of the University
of California, Los Angeles, and the Jet Propulsion Laboratory/California
Institute of Technology, funded by the National Aeronautics and
Space Administration.

The WISE data were all provided by the Infrared
Science Archive at Caltech.

\vspace{5mm}
\facilities{WISE}

\end{document}